\documentclass[aps,prb,twocolumn,superscriptaddress,showpacs,floatfix]{revtex4-1}

\usepackage{amsmath}
\usepackage{amssymb}
\usepackage{graphicx}
\usepackage{mathptmx} 
\usepackage{color}  

\begin{document}

\title{Charge density wave fluctuations in La$_{2-x}$Sr$_{x}$CuO$_{4}$ and their competition with superconductivity}

\author{T. P. Croft}
\affiliation{H. H. Wills Physics Laboratory, University of Bristol, Bristol, BS8 1TL, United Kingdom.}

\author{C. Lester}
\affiliation{H. H. Wills Physics Laboratory, University of Bristol, Bristol, BS8 1TL, United Kingdom.}

\author{M. S. Senn}
\affiliation{Diamond Light Source Ltd., Harwell Science and Innovation Campus, Didcot, Oxfordshire OX11 0DE, United Kingdom.}

\author{A. Bombardi}
\affiliation{Diamond Light Source Ltd., Harwell Science and Innovation Campus, Didcot, Oxfordshire OX11 0DE, United Kingdom.}

\author{S. M. Hayden}
\affiliation{H. H. Wills Physics Laboratory, University of Bristol, Bristol, BS8 1TL, United Kingdom.}

\begin{abstract}

We report hard (14 keV) x-ray diffraction measurements on three compositions ($x=0.11,0.12,0.13)$ of the high-temperature superconductor La$_{2-x}$Sr$_{x}$CuO$_{4}$. All samples show charge-density-wave (CDW) order with onset temperatures in the range 51--80~K and ordering wavevectors close to (0.23,0,0.5).  The CDW is strongest with the longest in-plane correlation length near 1/8 doping. On entering the superconducting state the CDW is suppressed, demonstrating the strong competition between the charge order and superconductivity. CDW order coexists with incommensurate magnetic order and the wavevectors of the two modulations have the simple relationship $\boldsymbol{\delta}_{\mathrm{charge}}= 2\boldsymbol{\delta}_{\mathrm{spin}}$. The intensity of the CDW Bragg peak tracks the intensity of the low-energy (quasi-elastic) spin fluctuations. We present a phase diagram of La$_{2-x}$Sr$_{x}$CuO$_{4}$ including the pseudogap phase, CDW and magnetic order.

\end{abstract}


\maketitle

\section{Introduction}

A large body of experimental evidence now suggests that charge density wave (CDW) order may be a generic feature of underdoped high-temperature cuprate superconductors \cite{Ghiringhelli2012_GLMB,Chang2012_CBHC,SilvaNeto2014_SAFC,Comin2014_CFYY,Wu2011_WMKH,Wu2013_WMKH,LeBoeuf2013_LKHL}. For example, in YBa$_2$Cu$_3$O$_{6+x}$, charge order has been detected in magnetic fields $\gtrsim 15$~T by NMR \cite{Wu2011_WMKH,Wu2013_WMKH} and ultrasound \cite{LeBoeuf2013_LKHL} indicating that it is essentially static. X-ray experiments \cite{Chang2012_CBHC,Ghiringhelli2012_GLMB} observe incommensurate charge density wave order in zero field which may only fluctuate on frequency scales less than $\sim$1~meV \cite{Blackburn2013_BCSL,LeTacon2014_LBSD}. Taken together, these experiments suggest that incommensurate charge correlations appear below the pseudogap temperature, compete with the superconductivity and become static in magnetic fields $\gtrsim 15$~T.

The La$_{2-x}$Sr$_{x}$CuO$_4$  (LSCO) system is a canonical example of high-$T_c$ superconductivity. It has a simple structure without the complications of the CuO chains and CuO$_2$ bilayers present in some other cuprates.  The LSCO system is interesting to study amongst the various cuprate superconductors because doping with Ba instead of Sr or adding Nd or Eu causes the material undergo a low temperature structural (LTT) phase transition not found elsewhere (see Sec.~\ref{Sec:Background}). This leads to charge ordering and the strong suppression or absence of superconductivity. The La$_{2-x}$Sr$_x$CuO$_4$ system can be thought of as a ``parent compound'' where the LTT transition does not occur.

Many physical properties of the LSCO system suggest the existence of charge ordering near doping $p \approx 1/8$. Firstly, the superconducting onset temperature $T_c$ as a function of doping shows a suppression of about 5~K, at $p \approx 1/8$, with respect to the general trend \cite{Kofu2009_KLFK,Yamada1998_YLKW}. This suggests the presence of a competing phase.   La$_{2-x}$Sr$_{x}$CuO$_4$ also shows a region of incommensurate spin density wave (SDW) order near 1/8 doping \cite{Wakimoto1999_WSEH,Lake2002_LRCA,Kofu2009_KLFK,Kimura1999_KHMY,Chang2008_CNGC,Julien2003_Juli}. The presence of SDW order follows charge ordering in closely related compounds such as La$_{2-x}$Ba$_x$CuO$_4$ (LBCO) \cite{Hucker2011_HZGX}.    NMR \cite{Hunt1999_HSTI, Mitrovic2008_MJVH} and Hall effect \cite{Hwang1994_HBTK} measurements have suggested that there may be charge order in La$_{2-x}$Sr$_{x}$CuO$_4$.  Extended
x-ray absorption fine structure (EXAFS) \cite{Bianconi1996_BSLM} and atomic pair distribution function (PDF) analysis of neutron scattering data \cite{Bozin1999_BBKT}  has provided evidence of ordering structural distortions.  Finally, the LSCO system \cite{Park2014_PFHL,McQueeney1999_MPEY} together with other cuprates \cite{Reznik2012_Rezn} show  (Kohn) anomalies in the optic phonons. These are often associated with charge ordering.

In this paper, we use 14~keV x-rays to observe charge density wave order in La$_{2-x}$Sr$_x$CuO$_4$. We studied three compositions of LSCO near 1/8 doping.  For the composition with the strongest CDW, La$_{1.78}$Sr$_{0.12}$CuO$_{4}$, the component of the ordering wavevector within the CuO$_2$ planes is $\mathbf{q}_{\mathrm{CDW}}=(0.235,0)$. This value is similar to that found in related compounds La$_{2-x}$Ba$_x$CuO$_4$ (LBCO), La$_{1.6-x}$Nd$_{0.4}$Sr$_x$CuO$_4$ (Nd-LSCO) and La$_{1.6-x}$Eu$_{0.4}$Sr$_x$CuO$_4$ (Eu-LSCO) which are either not superconducting or have suppressed superconducting $T_c$. In LSCO, we observe a suppression of the CDW on entering the superconducting state, demonstrating the strong competition between the charge order and superconductivity.

There have been three recent reports \cite{Wu2012_WBTC,Christensen2014_CCLF,Thampy2014_TDCI} of x-ray studies of the CDW in La$_{1.88}$Sr$_{0.12}$CuO$_4$, one of the compositions studied here, in the last two years. These studies used resonant and non-resonant diffraction techniques with x-ray energies from 529~eV  to 100~keV. The results presented here are broadly in agreement the very recent studies \cite{Christensen2014_CCLF, Thampy2014_TDCI}. The earlier study \cite{Wu2012_WBTC} concluded no bulk CDW was present. In the light of the results presented in this paper, we believe the authors probably arrived at this conclusion because they studied a CDW satellite position with small (or zero) structure factor.  We compare these studies with the present work in Sec.~\ref{sec:other_xray}.

\section{Background}\label{Sec:Background}

At high temperatures, La$_{2-x}$Sr$_x$CuO$_4$ has the so-called high temperature tetragonal (HTT) structure with space group $I4/mmm$, flat CuO$_2$ planes and lattice parameters $a=b \approx$3.78~\AA, $c\approx$13.2~\AA. We will use the lattice of this structure to describe real and reciprocal space in this paper.  The structure is built from copper-centered oxygen octahedra. Below $T_{\mathrm{LTO}} \approx240$~K, these rotate about the [110]-type directions to form a new low temperature orthorhombic (LTO, $Bmab$) structure.  The related materials LBCO, Eu-LSCO and Nd-LSCO exhibit an additional phase transition \cite{Axe1989_AMHC} to a low-temperature tetragonal phase (LTT, $P42/ncm$). This structure has octahedra rotated around [100]-type direction and appears to favor charge stripe formation. Thus these three materials all form stripe order at or below $T_{\mathrm{LTT}}$. However, no bulk LTT transition has been observed in La$_{2-x}$Sr$_x$CuO$_4$.

\begin{figure*}[ht]
\begin{center}
\includegraphics[width=0.95\linewidth]{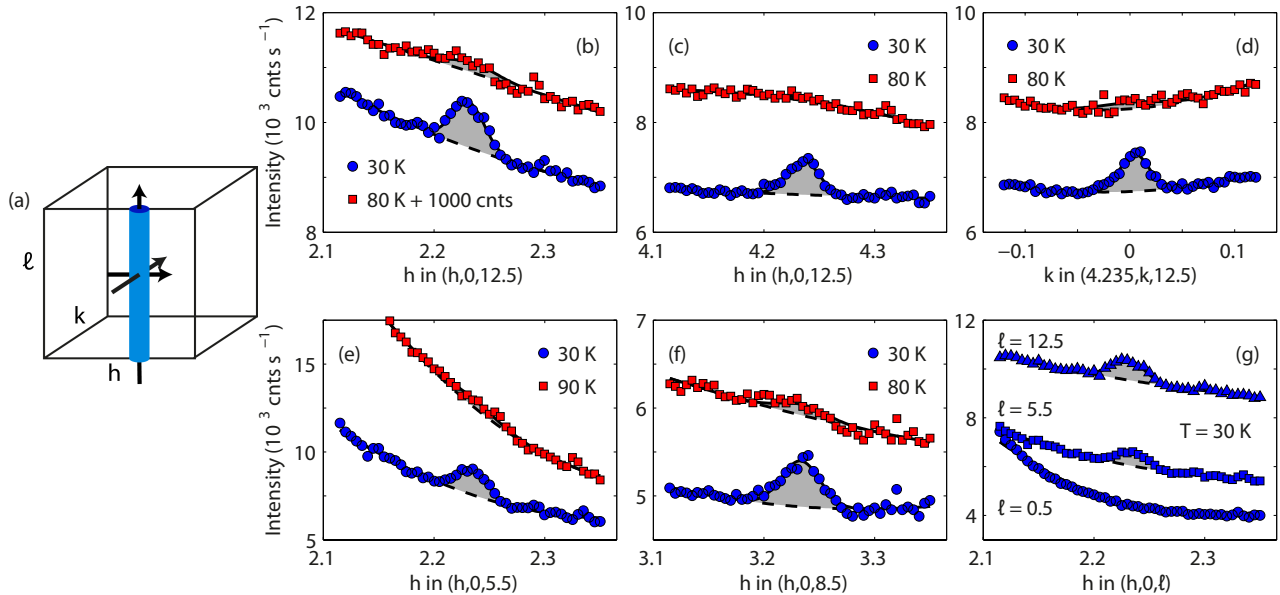}
\end{center}
\caption{(color online) (a) Schematic drawing of an intensity-modulated rod of scattering in reciprocal space due to the CDW. Trajectories of scans in other panels are shown. (b)--(f) CDW peaks for various $h$ and $\ell$ values for La$_{1.88}$Sr$_{0.12}$CuO$_4$.}.
\label{fig:strong_scans_LSCO12}
\end{figure*}

\begin{figure}[ht]
\begin{center}
\includegraphics[width=0.70\linewidth]{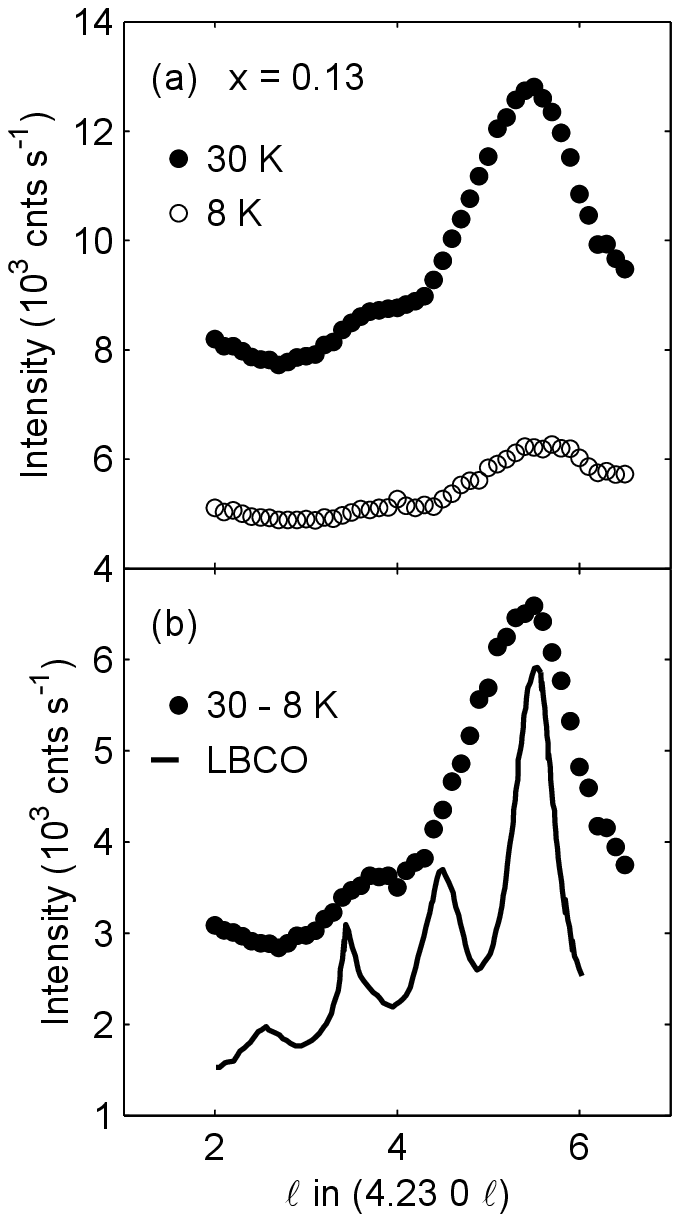}
\end{center}
\caption{(a) $\ell$-dependence of the scattering along the line (4+$\delta$,0,$\ell$) for $T=30$~K and $T=8$~K for La$_{1.87}$Sr$_{0.13}$CuO$_4$.  (b) Points show $\ell$-dependence of CDW scattering. The signal has been isolated by subtracting 8~K data, where the CDW signal is weak, from 30~K data where the signal is strongest. The solid line shows equivalent data collected on La$_{1.875}$Sr$_{0.125}$CuO$_4$ from Ref.~\onlinecite{Kim2008_KGGC}}.
\label{fig:l_scan_LSCO13}
\end{figure}

\section{Experimental Details}

\subsection{Samples}

Single crystals of La$_{2-x}$Sr$_x$CuO$_4$ with three compositions close to $x=1/8$ were grown by the travelling-solvent floating-zone technique using an infra-red image furnace. Further details of the growth method are given in Refs.~\onlinecite{Komiya2002_KASL,Lipscombe2009_LVPF}. Similar samples have been well characterized by inelastic neutron scattering (INS) \cite{Lipscombe2009_LVPF} and angle-resolved photoemission spectroscopy (ARPES) \cite{Chang2013_CMPC}.  The Sr stoichiometry, $x$, was measured by scanning electron microscopy with energy dispersive x-ray analysis (EDX) and also by inductively-coupled plasma atomic-emission spectroscopy (ICP-AES). Superconducting transition temperatures ($T_c$) were determined using a Quantum Design MPMS magnetometer with samples cooled in a 1~Oe field. The results of these characterizations are shown in Table~\ref{table:character}. In order to carry out x-ray experiments, samples were cut into plates with a (100) face and typical dimensions $2 \times 3 \times 0.5$~mm. The plate faces were polished to 0.3 $\mu$m and etched in 0.03M HCl. The samples were then annealed in oxygen at 800$^{\circ}$C.

\subsection{x-ray experiments}

X-ray diffraction (XRD) experiments were performed on the I16 beam line at the Diamond Light Source (DLS).  The sample was mounted in a closed-cycle cryostat on a six-circle kappa diffractometer.  We used a vertical scattering geometry. Experiments were performed in reflection geometry with 14~keV x-rays which have a penetration depth of 21~$\mu$m. Data were collected in bisecting-mode to reduce absorption corrections. In this mode the angle of incidence and angle of refection of the x-rays are equal. We label reciprocal space $(h,k,\ell)$ in units of ($2\pi/a$,$2\pi/b$,$2\pi/c$) of the HTT structure of La$_{2-x}$Sr$_x$CuO$_4$. Our samples become twinned below the LTO phase transition which occurs at $T_{\mathrm{LTO}} \approx 240$~K for the present compositions. Below $T_{\mathrm{LTO}}$, we do not distinguish between the orthorhombic $a$ and $b$ axes in this paper.

\begin{table}[htb]
\begin{center}
\begin{ruledtabular}
\begin{tabular}{ccccc}
 $x$ in La$_{2-x}$Sr$_x$CuO$_4$ & $T_{\mathrm{c}}~$(K)  & $T_{\mathrm{CDW}}$~(K)
 & $\delta$ (r.l.u.) & $\xi_a(T=T_c)$ (\AA) \\ \hline
 0.110(2)& 24.4(2) & 51(5) &  0.224(3) & 19(4) \\
 0.120(2)& 29.5(2) & 75(10) & 0.235(3) & 30(4) \\
 0.130(2)& 30.4(2) & 80(20) & 0.232(3) & 25(4) \\
\end{tabular}
\end{ruledtabular}
\end{center}
\caption{Characteristics of the La$_{2-x}$Sr$_{x}$CuO$_4$ samples studied. Composition ($x$) evaluated from average of EDX and ICP-AES.  The superconducting onset $T_{\mathrm{c}}$ was determined from the 1 Oe field-cooled magnetization. CDW order has onset temperature $T_{\mathrm{CDW}}$ and ordering wavevector $(\delta,0,0.5)$. The CDW correlation length in the $\mathbf{a}$-direction is $\xi_a$.}
\label{table:character}
\end{table}

\section{Results}

A charge density wave gives rise to a modulation of the atomic positions throughout the crystal resulting in satellite peaks at reciprocal space positions $\mathbf{Q}=\boldsymbol{\tau}+\mathbf{q}_{\mathrm{CDW}}$, where $\boldsymbol{\tau}$ are reciprocal lattice positions of the unmodulated structure and $\mathbf{q}_{\mathrm{CDW}}$ is the wavevector of the CDW.  X-ray experiments \cite{Ghiringhelli2012_GLMB,Chang2012_CBHC,Blackburn2013_BCHH} on YBCO showed that certain reciprocal lattice positions of the unmodulated structure were surrounded by satellite peaks with wavevectors $\mathbf{q}_{\mathrm{CDW}}=(\pm \delta, 0, 1/2)$ and $(0, \pm \delta, 1/2)$. To be more precise, the peaks are actually ``rods'' of scattering in reciprocal space which are parallel to $\mathbf{c}^{\star}$ as shown in Fig.~\ref{fig:strong_scans_LSCO12}(a). The intensity of the rods is modulated as a function of $\ell$ ($\mathbf{q}=\ell \mathbf{c}^{\star}$) with peaks at half-integer positions \cite{Chang2012_CBHC,Blackburn2013_BCHH}. It has also been found \cite{Kim2008_KGGC,Hucker2011_HZGX} that LBCO exhibits charge ordering peaks which are strongest at half-integer positions in $\ell$. In La$_{2-x}$Sr$_x$CuO$_4$, a previous x-ray study \cite{Wu2012_WBTC} reports that near-surface scattering gives rise to CDW peaks with $\mathbf{q}_{\mathrm{CDW}}=(0.24,0,0)$ below 55~K.

\begin{figure}[t]
\begin{center}
\includegraphics[width=0.95\linewidth]{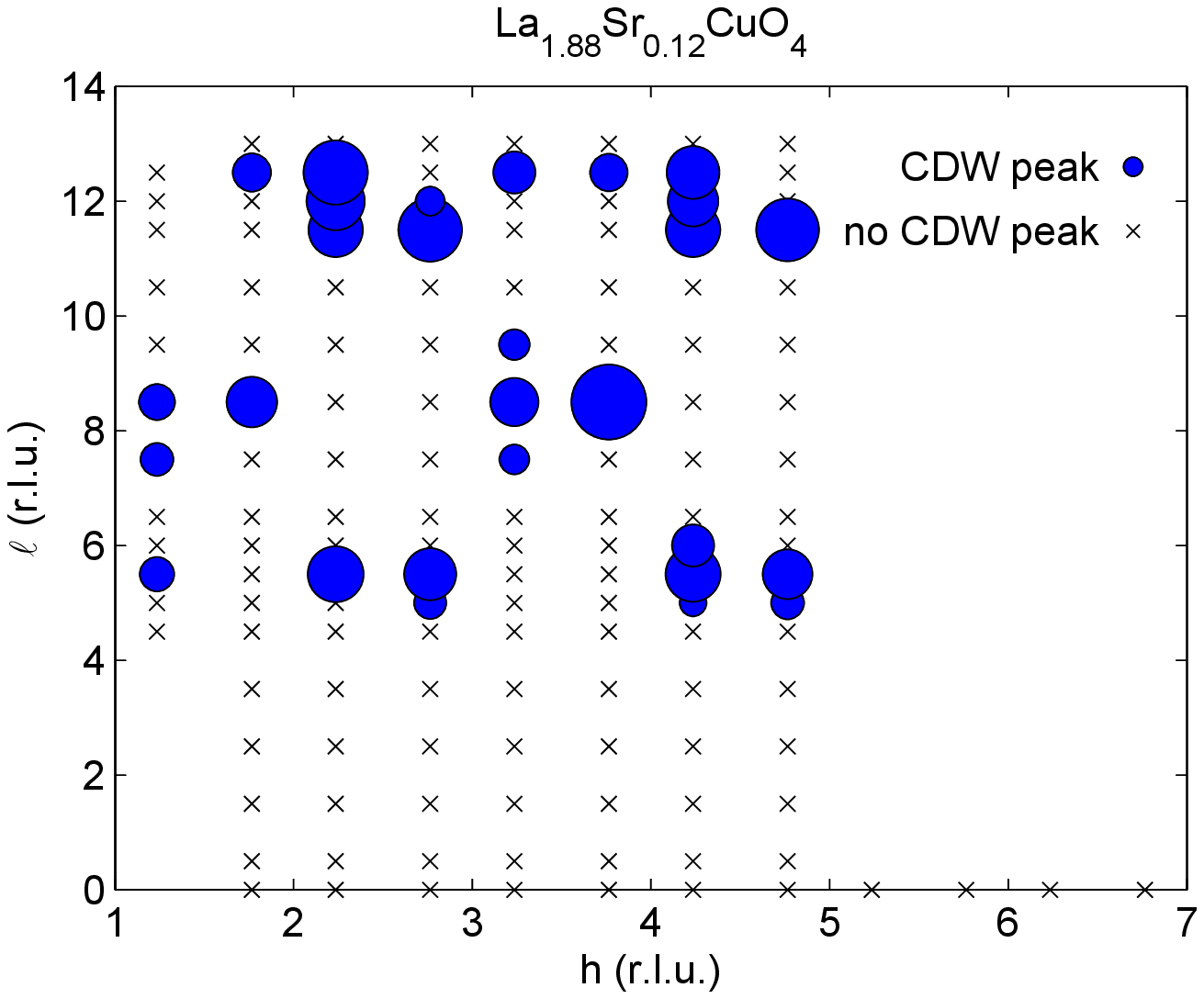}
\end{center}
\caption{(color online) Intensities of CDW satellite peaks in the $(h,0,\ell)$ plane of reciprocal space. The area of the filled circles are proportional to the CDW peak intensities. Crosses ($\times$) are positions investigated where no CDW peak was detected.}
\label{fig:LSCO12_0kl_0q0_map}
\end{figure}

We first describe the scattering observed from our $x=0.12$ sample. Fig.~\ref{fig:strong_scans_LSCO12} shows $h$-scans (parallel to $\mathbf{a}^{\star}$) near various reciprocal lattice positions. Data were collected at $T=30$~K~$\approx T_c$ (were CDW scattering in YBCO was found to be strongest) and at $T=80-90$~K as a background.  Following previous work on YBCO and LBCO, we made scans at half-integer positions in $\ell$. Fig.~\ref{fig:strong_scans_LSCO12}(g) illustrates the $\ell$-dependence of the CDW scattering showing scans through $(2+\delta,0,\ell)$ for some characteristic $\ell$ positions. We were unable to observe CDW peaks for $0 \leq \ell \leq 4.5$. For example, Fig.~\ref{fig:strong_scans_LSCO12}(g) shows a scan with $\ell=0.5$ which has no discernible incommensurate peak. In contrast, CDW peaks are observed at $\ell=5.5$ and $\ell=12.5$. Strong CDW peaks are also observed for $\ell=12.5$ when we scan through the $(4+\delta,0,\ell)$ position as shown in Fig.~\ref{fig:strong_scans_LSCO12}(c). For the $(3+\delta,0,\ell)$ position, the peaks are strongest near $\ell=8.5$ as shown in Fig.~\ref{fig:strong_scans_LSCO12}(f). Fig.~\ref{fig:LSCO12_0kl_0q0_map} summarizes the measured peak intensities in the $(h,0,\ell)$ plane of reciprocal space. On increasing the temperature to 80~K the CDW peaks are largely suppressed. However, there may be a weak peak near $\mathbf{q}_{\mathrm{CDW}}$ at 80~K and above (see for example data at 80~K in panels (b) and (f)). We return to this point below.

By fitting Gaussian curves to the data in Fig.~\ref{fig:strong_scans_LSCO12}, we can estimate the correlation length $\xi$ of the charge order from the Gaussian width parameter $\sigma$ as $\xi=1/\sigma$. For our $x=0.12$ sample, we find that the in-plane correlation lengths parallel and perpendicular to $\mathbf{q}_{\mathrm{CDW}}$ are $\xi_{\parallel}= 30 \pm 4$~\AA\ and $\xi_{\perp}= 31 \pm 4$~\AA\ at 30~K.  The data (circles) in Fig.~\ref{fig:l_scan_LSCO13}(b) show the $\ell$-dependence of the CDW intensity. From the width of the peak near $\ell=5.5$, we estimate the correlation length along the $c$-axis as $\xi_c=3.5 \pm 0.5$~\AA\ for the $x=0.13$ sample. This corresponds to about half of the separation of the CuO$_2$ planes ($\approx6.6$~\AA).

\begin{figure}[t]
\begin{center}
\includegraphics[width=0.75\linewidth]{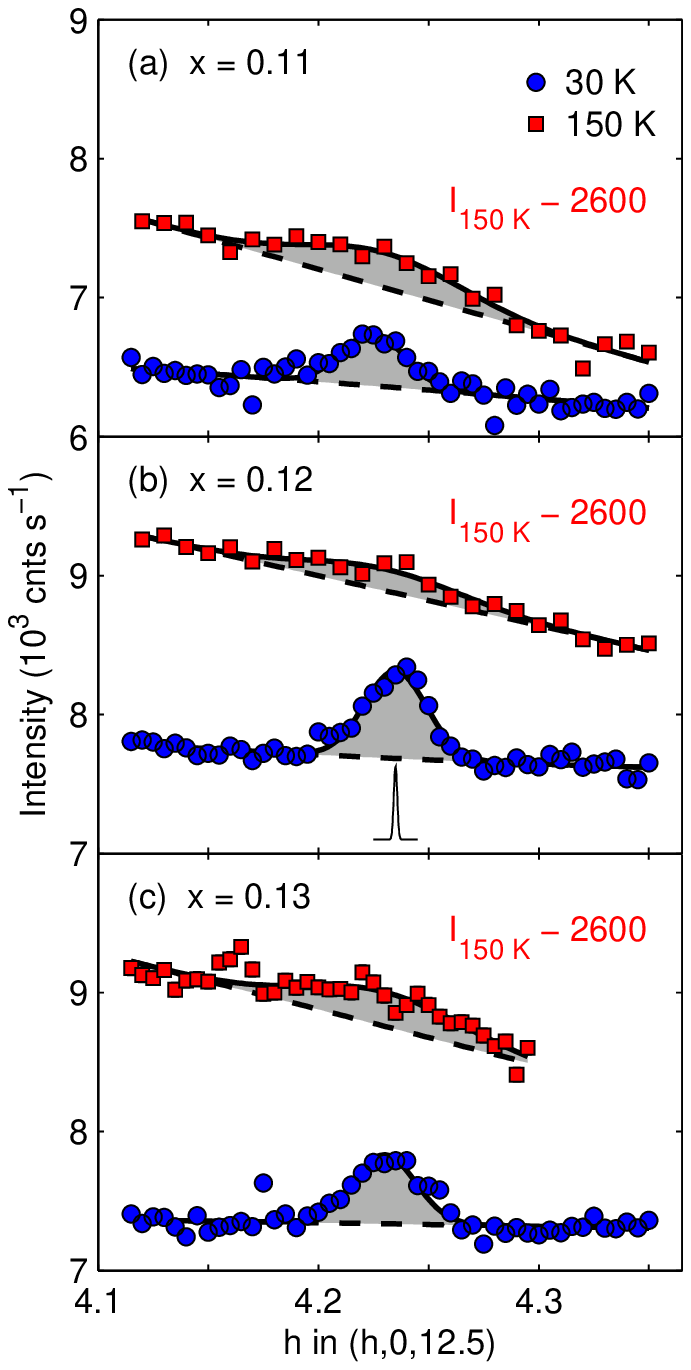}
\end{center}
\caption{ (color online) (a)-(c) Scans through the CDW satellite peak at $T$=30~K$~\gtrsim T_c$ and at 150~K showing the doping dependence of the incommensurate wavevector and approximate strength of the peak. The high temperature scans have been offset for clarity. Solid line in (b) shows the instrumental resolution. }
\label{fig:incommensurability_scans}
\end{figure}

\begin{figure}[htb]
\begin{center}
\includegraphics[width=0.8\linewidth]{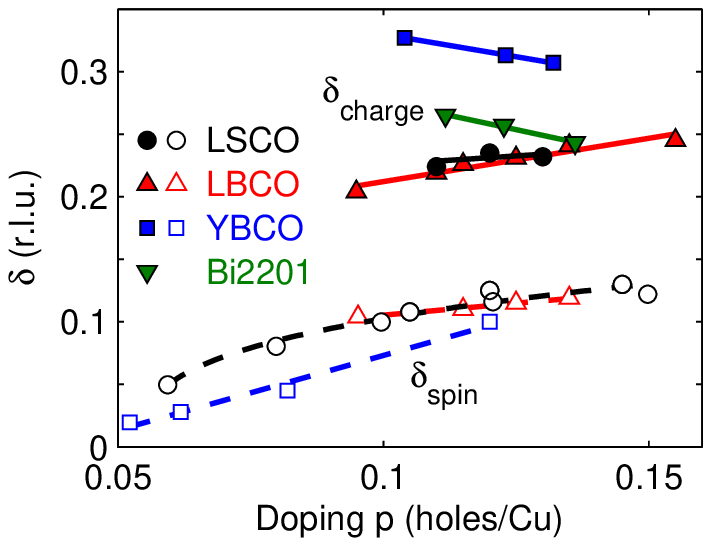}
\end{center}
\caption{ (color online) Spin and charge incommensurability versus doping for La$_{2-x}$Sr$_{x}$CuO$_4$ (LSCO) \cite{Wakimoto1999_WSEH,Chang2008_CNGC}, La$_{2-x}$Ba$_{x}$CuO$_4$ (LBCO) \cite{Hucker2011_HZGX}, YBa$_2$Cu$_3$O$_{6+x}$ \cite{Haug2010_HHSB}, and
Bi$_{2}$Sr$_{2-x}$La$_{x}$CuO$_{6+x}$ (Bi2201) \cite{Comin2014_CFYY}. Open symbols represent spin order or the strong spin fluctuations with in-plane wavevector $(1/2+\delta_{\mathrm{spin}},1/2)$. Closed symbols are CDW peaks at $(\delta_{\mathrm{charge}},0)$
}
\label{fig:incommensurability}
\end{figure}

In order to determine the doping dependence of the incommensurability, amplitude and onset temperature of the CDW, we studied three compositions. Fig.~\ref{fig:incommensurability_scans} shows scans measured near $T_c$ and high temperature backgrounds.  These scans were collected with the same spectrometer conditions and with samples of similar geometry. Thus the scattering intensities should be directly comparable. We find that the relative heights of the $(4+\delta,0,12.5)$ CDW peaks are 160, 430 and 340 with respect to the high-temperature background for $x=0.11, 0.12, 0.13$. Thus the CDW in the $x=0.11$ sample is notably weaker than the other compositions.  The incommensurability parameters $\delta$ and in-plane correlation lengths are given in Table~\ref{table:character} and plotted in Fig.~\ref{fig:incommensurability}.

\begin{figure}[htb]
\begin{center}
\includegraphics[width=0.95\linewidth]{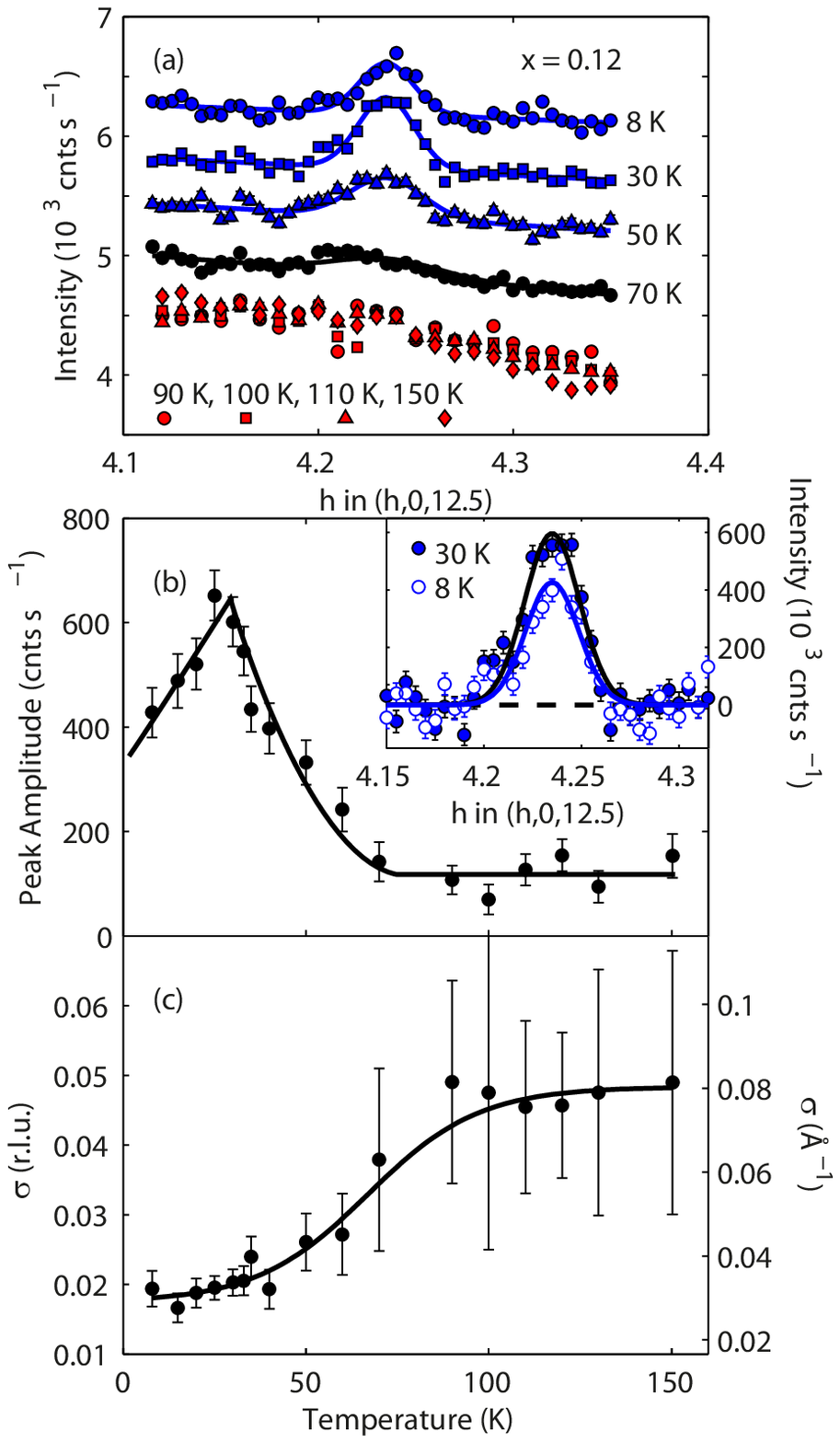}
\end{center}
\caption{(color online) (a) $h$-dependent scans through the $(\delta,0,12.5)$ CDW peak for various temperatures for La$_{1.88}$Sr$_{0.12}$CuO$_4$.  Solid lines are fits to a Gaussian lineshape. All scans, except for $T=8$~K, have been offset for clarity. (b),(c) Peak heights and widths of the CDW peak extracted from the fits in panel (a). The inset to (b) shows 30~K and 8~K data from (a) plotted together with linear backgrounds subtracted. This illustrates the suppression of the CDW in the superconducting state.}
\label{fig:temp_dep_0_k_12p5_LSCO12}
\end{figure}

\begin{figure}[tb]
\begin{center}
\includegraphics[width=0.95\linewidth]{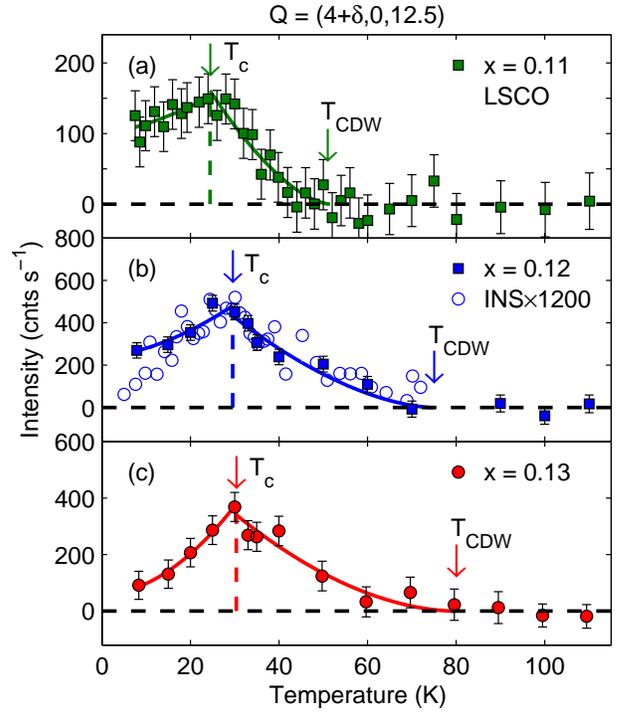}
\end{center}
\caption{(color online) Temperature dependence of the peak intensity of the $(4+\delta,0,12.5)$ CDW peak for Sr dopings $x$=0.11,0.12 and 0.13.  Peak heights are determined from fitting data such as that displayed in Fig.~\ref{fig:temp_dep_0_k_12p5_LSCO12}. Heights are plotted with respect to average height in the range 90--150~K. The solid lines are power law fits described in the main text. The open circles in panel (b) show inelastic neutron scattering (INS) measurements of spin fluctuations at the spin ordering wavevector $\mathbf{q}_{\mathrm{SDW}}=(1/2+\delta_{\mathrm{spin}},1/2)=(0.625,0.5)$ and energy transfer $E=0.3$~meV for La$_{1.88}$Ba$_{0.12}$CuO$_4$ from Ref.~\onlinecite{Romer2013_RCCA}. The intensities of the INS data have been multiplied by 1200.}
\label{fig:temp_dep_CDW}
\end{figure}

Fig.~\ref{fig:temp_dep_0_k_12p5_LSCO12}(a) shows a series of $h$-scans through the (4+$\delta$,0,12.5) position for temperatures between 8~K and 150~K.  Inspection of the data suggests that, at high temperature $T \geq 70$~K, there is a peak (on a sloping background) near $h=4.235$.  The peak's height is approximately independent of temperature above 70~K. Below about $T=70$~K, a stronger peak develops. This may be because of the appearance of a second component or an evolution of the original peak. The peak height increases until $T=33$~K~$\sim T_c$ and then decreases.  Fitting the scans to a Gaussian lineshape yields the temperature dependence of the peak amplitude and width shown in \ref{fig:temp_dep_0_k_12p5_LSCO12}(b),(c). The origin of the high temperature peak is unclear (see Sec.~\ref{sec:charge_order}). However it is clear that the peak amplitude has two components which can be phenomenologically separated. One of the components is weaker, broader in  $\mathbf{q}$ and either appears above 150~K (the highest temperature measured) or is always present. The second component is sharper, appears at about 70~K, and gains strength on the approach to $T_c$. Subtracting the broad high temperature component measured for $90 \leq T \leq 150$~K, we obtain the temperature dependence of the amplitude of the low temperature component shown in Fig.~\ref{fig:temp_dep_CDW}. Note that the samples with $x=0.11$ and $x=0.13$ also show a weak high temperature component. We associate the onset of the stronger component with a CDW transition, hence we label its onset temperature as $T_{\mathrm{CDW}}$.

\begin{figure}[htb]
\begin{center}
\includegraphics[width=0.95\linewidth]{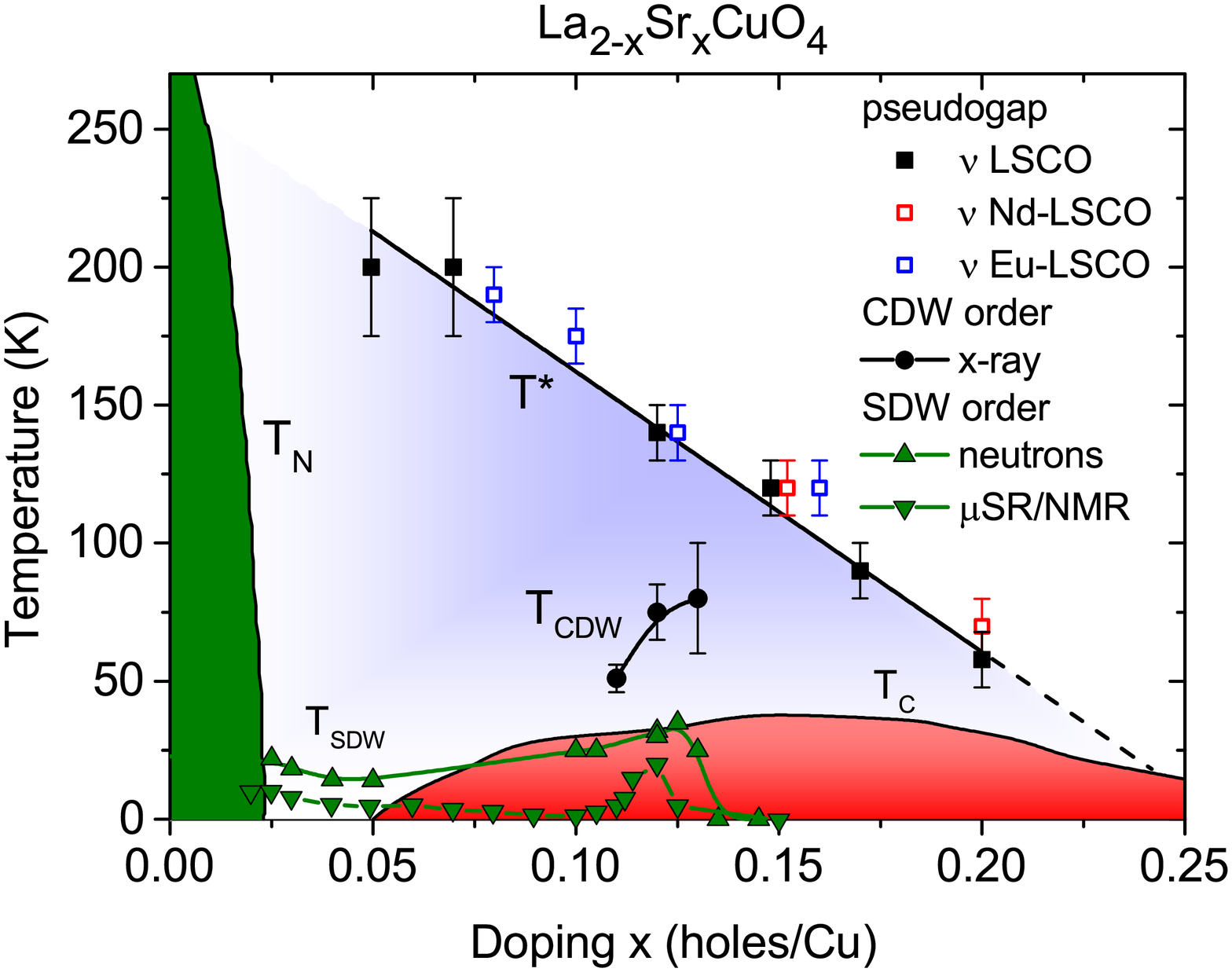}
\end{center}
\caption{(color online) Temperature versus doping phase diagram of La$_{2-x}$Sr$_{x}$CuO$_4$. $T_{\mathrm{CDW}}$ is the onset temperature of charge-density-wave order determined from the present x-ray experiment. $T_{\mathrm{SDW}}$ is the onset temperature of the incommensurate magnetic order observed with neutron scattering \cite{Wakimoto1999_WSEH,Lake2002_LRCA,Kofu2009_KLFK,Kimura1999_KHMY,Chang2008_CNGC},  nuclear magnetic resonance (NMR) and muon spin resonance ($\mu$SR) \cite{Julien2003_Juli}.
$T_c$ is the superconducting transition temperature from Ref.~\onlinecite{Yamada1998_YLKW}. $T^{\star}$ is the pseudogap onset temperature determined from the upturn in the Nernst coefficient \cite{Doiron-Leyraud2012_DoTaa,Wang2006_WaLO}.
}
\label{fig:phase_diagram}
\end{figure}

\section{Discussion}\label{sec:Discussion}

\subsection{Nature of charge order}
\label{sec:charge_order}

Charge density wave order has now been observed near 1/8 doping in superconducting cuprates by various types of x-ray diffraction in YBa$_2$Cu$_3$O$_{6+x}$ \cite{Chang2012_CBHC,Ghiringhelli2012_GLMB}, Bi$_{2}$Sr$_{2}$CaCu$_{2}$O$_{8+x}$ (Bi2212) \cite{SilvaNeto2014_SAFC} and Bi$_{2}$Sr$_{2-x}$La$_{x}$CuO$_{6+x}$ (Bi2201) \cite{Comin2014_CFYY} and also in a number of related weakly or non-superconducting cuprates including La$_{2-x}$Ba$_{x}$CuO$_{4}$ \cite{Kim2008_KGGC,Hucker2011_HZGX} and La$_{1.8-x}$Eu$_{0.2}$Sr$_{x}$CuO$_{4}$ (Eu-LSCO) \cite{Fink2009_FSWG}. The onset of CDW order in YBCO in zero magnetic field is accompanied by a downturn in the Hall coefficient \cite{LeBoeuf2013_LKHL} signalling electronic reconstruction. A similar anomaly is observed at the CDW transition in Eu-LSCO \cite{Taillefer2010_Tail}. YBCO also shows the onset of a Kerr effect \cite{Xia2008_XSDK} at the CDW transition.  The CDW can be also be detected in YBCO through the modification of the NMR lineshape \cite{Wu2011_WMKH} and through the change of elastic constants seen in ultrasound \cite{LeBoeuf2013_LKHL}.  However, NMR and ultrasound only detect the CDW at finite magnetic field $B \gtrsim 15$~T and $T \lesssim 70$~K. This suggests that the state detected by x-ray scattering is actually still fluctuating and that the application of a large magnetic field can cause it to lock to the crystal lattice. Throughout this discussion, we will use the term ``CDW'' to describe the state observed by x-rays unless otherwise stated.

In the case of La$_{2-x}$Sr$_{x}$CuO$_{4}$ ($x=0.125$), the Hall coefficient $R_H$ shows a downturn \cite{Hwang1994_HBTK} at $T_H \approx$70~K consistent with appearance of a CDW. The NMR ${}^{139}$La linewidth \cite{Mitrovic2008_MJVH} also broadens below 80~K for $x=0.12$. We therefore conclude that the component of the signal shown in Fig.~\ref{fig:temp_dep_CDW} is due to the appearance of a CDW at $T_{\mathrm{CDW}}$. A possible origin of the weak residual peaks observed for $T > T_{\mathrm{CDW}}$ in Figs.~\ref{fig:strong_scans_LSCO12},\ref{fig:incommensurability_scans},\ref{fig:temp_dep_0_k_12p5_LSCO12}(a) is the presence of local regions of the sample with the low-temperature tetragonal (LTT) structure \cite{Axe1989_AMHC}. Such regions have been identified in similar samples from atomic pair distribution function (PDF) analysis of neutron powder-diffraction data \cite{Bozin1999_BBKT} and also in transmission electron microscopy \cite{Horibe2000_HoIK}. The CuO$_6$ octahedron rotation around the crystallographic [100] axis associated with the LTT phase favors charge ordering with a wavevector close to the one reported here. Indeed, LBCO \cite{Fujita2004_FGYT,Hucker2011_HZGX} and Nd-LSCO \cite{Tranquada1995_TSAN} have CDW or stripe-order transitions concomitant with their LTO-LTT structural phase transitions. Defect regions or twin boundaries with a local LTT structure within a mainly LTO twinned crystal of LSCO would locally seed and pin a region with CDW order. Such regions might exist up to $T_{\mathrm{LTO}} \approx 240$~K in our samples.

Non-resonant x-ray diffraction such as the experiments performed here are primarily sensitive to the atomic displacements. Further, the intensity of satellite peaks $I \propto (\boldsymbol{\varepsilon} \cdot \mathbf{Q})^2$, where  $\boldsymbol{\varepsilon}$ is the displacement of the atoms.  The fact that we do not see strong CDW peaks for small $\ell$ values (See Fig.~\ref{fig:LSCO12_0kl_0q0_map}) suggests that the atomic displacements associated with the CDW in LSCO have a large $\mathbf{c}$ component. This is also the case in YBCO \cite{Blackburn2013_BCHH}.  Presumably the $\mathbf{c}$ motion is associated with the tilting or ``breathing'' of the CuO$_6$ octahedra and the concomitant displacement of the large Z atoms - La and Cu.

\subsection{Incommensurability, correlation lengths and temperature dependence}

Fig.~\ref{fig:incommensurability} shows the incommensurability, $\delta$, of the CDW plotted against doping compared with a number of other systems.  We note that there is little change in $\delta$ over the range of doping investigated in the present experiment. Our data are consistent with the trend line of LBCO \cite{Hucker2011_HZGX}, with $\delta$ increasing with doping.  In contrast, YBCO \cite{Blackburn2013_BCHH} and Bi2201 \cite{Comin2014_CFYY} show incommensurabilities (see Fig.~\ref{fig:incommensurability}) that decrease with increasing doping and have higher values for the same doping level. Authors of Ref.~\onlinecite{Comin2014_CFYY} propose that wavevector of the CDW is determined by anti-nodal nesting at Fermi surface hot spots. It is unclear whether the different trend seen in LSCO can be explained by the same mechanism.

The anomalous dispersion of phonons can also be used to detect anomalies in the charge response. It has been known for some time that the optic phonons in LSCO \cite{McQueeney1999_MPEY,Park2014_PFHL}  and other cuprates \cite{Reznik2012_Rezn} show such anomalies. In La$_{1.85}$Sr$_{0.15}$CuO$_4$ anomalies are observed \cite{McQueeney1999_MPEY} at $\mathbf{q}=(0.25,0,0)$, i.e. with $\delta=0.25$~r.l.u. which is consistent with the trend line of Fig.~\ref{fig:incommensurability} for the charge peaks in LSCO and LBCO. Thus it appears that the wavevectors of the phonon anomalies and the charge ordering peaks observed here have a common origin. We would also expect further temperature-dependent anomalies in the acoustic phonons as recently observed \cite{Blackburn2013_BCSL,LeTacon2014_LBSD} in YBCO.

It is widely believed that the spin and charge correlations in cuprates are closely related.  In a simple stripe picture of intertwined spin and charge correlations \cite{Zaanen1989_ZaGu, Machida1989_Mach, Kivelson2003_KBFO}, the underlying antiferromagnetism (AF) and charge density have modulations characterized by wavevectors $\boldsymbol{\delta}_{\mathrm{spin}}$ and $\boldsymbol{\delta}_{\mathrm{charge}}$ respectively.  These yield spin and charge peaks at positions $\tau_{\mathrm{AF}} \pm \boldsymbol{\delta}_{\mathrm{spin}}$ and
$\tau_{\mathrm{lattice}} \pm \boldsymbol{\delta}_{\mathrm{charge}}$, where $\boldsymbol{\delta}_{\mathrm{charge}}= 2\boldsymbol{\delta}_{\mathrm{spin}}$.
This simple relationship describes observations in LBCO \cite{Hucker2011_HZGX} (see Fig.~\ref{fig:incommensurability}), Nd-LSCO  \cite{Tranquada1995_TSAN} and also in chromium \cite{Fawcett1988_Fawc}. In contrast, this relationship seems to break down in YBCO suggesting that the spin and charge correlations are not so directly connected.

The width of our CDW peaks yields the correlation length $(\xi=1/\sigma)$ of the CDW. In common with other superconducting cuprates, we find a relatively short in-plane correlation length with $\xi_a(T=T_c)= 30 \pm 4$~\AA. This compares with  $\xi_a (T=T_c) \approx$70~\AA\ for YBCO \cite{Chang2012_CBHC} and $\xi_a(T=T_c) \approx 20-30$~\AA\ in Bi2201. Thus in all these cases the CDW does not form a long range ordered state. This is possibly because the CDW is inherently fluctuating and in competition with superconductivity even above $T_c$ \cite{Hayward2014_HHMS}.  The CDW in LSCO is very weakly correlated along $\mathbf{c}$ with $\xi_c(T=T_c)=3.5 \pm 0.5$~\AA.

Fig.~\ref{fig:temp_dep_CDW}(a)--(c) show the temperature dependence of the CDW amplitude for the three compositions. A number of interesting features can be noted. As mentioned earlier, the CDW appears to be strongest for $x=0.12$. All the curves exhibit a concave upwards shape to the temperature dependence of the height above $T_c$ (i.e. $I \propto (T_{\mathrm{CDW}}-T)^{\beta}$ with {$\beta= 1.6-1.9 > 1$}).  This behavior is also observed in YBCO \cite{Chang2012_CBHC,Ghiringhelli2012_GLMB} and is probably a consequence of the fluctuating nature of the CDW observed (i.e. we are not observing a `true' phase transition). This picture is supported by recent theory \cite{Hayward2014_HHMS} in which superconducting and charge-density wave orders exhibit angular fluctuations in a six-dimensional space. As the superconductivity sets in at $T_c$, the CDW is suppressed. The $x=0.13$ sample has the highest $T_c$ and is closest to optimal doping. It shows the strongest suppression, with superconductivity almost ejecting the CDW at $T=8$~K.

\subsection{Phase diagram}

In Fig.~\ref{fig:phase_diagram} we combine our results with those from some other techniques to propose a phase diagram for La$_{2-x}$Sr$_{x}$CuO$_{4}$. An important boundary is that of the pseudogap phase $T^{\star}(x)$ which in LSCO can be identified from an upturn in the Nernst coefficient \cite{Doiron-Leyraud2012_DoTaa,Wang2006_WaLO}. From Fig.~\ref{fig:phase_diagram} we see that, as in the case of YBCO \cite{Chang2012_CBHC}, CDW order develops within the pseudogap phase.

LSCO develops incommensurate (IC) low-frequency magnetic correlations or spin-density wave (SDW) quasistatic order \cite{Wakimoto1999_WSEH,Lake2002_LRCA,Kofu2009_KLFK,Kimura1999_KHMY,Chang2008_CNGC,Julien2003_Juli} for a range of dopings $0.06 \lesssim x \lesssim 0.135$ at $\mathbf{q}_{\mathrm{SDW}}$. More precisely, there is a component of the spin-fluctuation spectrum, $|m (\mathbf{q}_{\mathrm{SDW}},\omega)|^2$, which is centered on $\omega=0$ with a temperature-dependent intensity and energy-width ( $\hbar \Gamma$). Because $\Gamma$ increases with temperature, the onset temperature $T_{\mathrm{SDW}}$ at which the SDW order can be detected depends on the frequency or frequency resolution of the measurement probe. When sufficient spectral weight is present in the frequency window of the probe, `order' is observed. For $\mu$SR and NMR the relevant energies (frequencies) are in the range $0.01-1$~$\mu$eV. These probes \cite{Julien2003_Juli} yield the lower line for $T_{\mathrm{SDW}}$ in Fig.~\ref{fig:phase_diagram}. The quasistatic order is also observed with cold-neutron scattering \cite{Wakimoto1999_WSEH,Lake2002_LRCA,Kofu2009_KLFK,Kimura1999_KHMY,Chang2008_CNGC}, this case the energy resolution is several orders of magnitude larger $\sim 0.2$~meV and a higher onset temperature is observed.

We note that for $\mu$SR, NMR, and neutron scattering, the onset temperature of the SDW in LSCO is enhanced near $x \approx 1/8$, where CDW order is observed. In this region of the phase diagram, the wavevectors of the two types of correlation have the simple relationship $\boldsymbol{\delta}_{\mathrm{charge}}= 2\boldsymbol{\delta}_{\mathrm{spin}}$ suggesting that the two types of order are intertwined. We further highlight this connection by considering the onset temperature for the SDW order measured on a higher frequency scale. In Fig.~\ref{fig:temp_dep_CDW}(b), we plot the inelastic neutron scattering measurements of the intensity of the magnetic fluctuations from Ref.~\onlinecite{Romer2013_RCCA} for a similar sample and for $\hbar \omega=0.3$~meV and $\mathbf{q}=\mathbf{q}_{\mathrm{SDW}}$. The x-ray and neutron intensities track each other, even to the extent that both are suppressed on entering the superconducting state. One should note that our x-ray measurements are collected without energy analysis the therefore with a large ($\gg 1$~meV) integration in frequency.

\subsection{Comparison with La$_{2-x}$Ba$_{x}$CuO$_{4}$}

It is interesting to compare La$_{2-x}$Sr$_{x}$CuO$_{4}$ with its sister system La$_{2-x}$Ba$_{x}$CuO$_{4}$. One of the major differences between the two systems is the strong suppression of superconductivity in LBCO, with $T_c$ being suppressed to 3~K \cite{Hucker2011_HZGX} for doping $p$=1/8, compared to $T_c \sim$~30K for LSCO of the same composition.  The LSCO and LBCO systems share the same HTT structure at high temperatures. However, LBCO undergoes an additional phase transition to a LTT structure at $T_{\mathrm{LTT}} \approx 54$~K. CDW order appears at this transition in this system. The charge order in LBCO is characterized by larger correlation lengths along $\mathbf{a}$ and $\mathbf{c}$, of $\xi_a \approx 125$~\AA\ and $\xi_c \approx 9$~\AA\ for $x=1/8$.  These compare with $\xi_a \approx 30$~\AA\ and $\xi_c \approx 3.5$~\AA\ for the $x=0.13$ sample studied here. The difference between the correlations along $\mathbf{c}^{\star}$ can be seen in Fig.~\ref{fig:l_scan_LSCO13}. In LBCO, Li \textit{et al.} \cite{Li2007_LHGT} have found the charge ordering transition coincides with the beginning of a rapid increase in the anisotropy of the resistivity between the CuO$_2$ planes and the $\mathbf{c}$-axis. This suggests that the dominant impact of the ordering is to electronically decouple the CuO$_2$ planes leading to 2D superconductivity \cite{Berg2007_BFKK,Li2007_LHGT} and the frustration of 3D superconducting phase order. In contrast, the less developed charge order in LSCO means that 3D superconductivity with a higher onset temperature is allowed to develop.

\subsection{Comparison with other x-ray studies}
\label{sec:other_xray}

As mentioned in the introduction there have been three other x-ray studies of the CDW in La$_{1.88}$Sr$_{0.12}$CuO$_{4}$ in the past two years. In this section, we compare our results with these studies. The studies were carried out with resonant diffraction at the copper $L_3$ \cite{Wu2012_WBTC,Thampy2014_TDCI} and oxygen $K$ \cite{Wu2012_WBTC} edges at energies 931~eV and 529~eV and also with non-resonant diffraction with energies of 8.9~keV \cite{Thampy2014_TDCI}, 14~keV (this study) and 100~keV \cite{Wu2012_WBTC,Christensen2014_CCLF}.  Resonant x-ray diffraction (RXRD) is sensitive to one atom type in the structure, the atomic scattering factor depends on the local electronic structure of the atom investigated. In the presence of a CDW the atomic scattering factor will be modulated \cite{Abbamonte2006_Abba} as the local environment and the valance of the atom are modulated in space. For a single atom type, the intensity of the satellite peaks in non-resonant x-ray diffraction (NRXRD) is $I \propto (\boldsymbol{\varepsilon} \cdot \mathbf{Q})^2$ (see Sec.~\ref{sec:charge_order}). Another difference between the various x-ray set ups is the penetration depth of the x-rays which generally increases with energy, but decreases in the locality of an absorption edge. Thus the Cu-RXRD and 14~keV experiments probe 0.1~$\mu$m and 10~$\mu$m into the sample respectively, while the 100~keV experiments probe the whole sample in transmission.

The RXRD studies at the O-$K$ \cite{Wu2012_WBTC} and Cu-$L$ \cite{Wu2012_WBTC,Thampy2014_TDCI} edges all report CDW order. Although Ref.~\onlinecite{Wu2012_WBTC} attributed their observations to the presence of a CDW at the surface. The first 100~keV study \cite{Wu2012_WBTC} did not observe CDW order. This is most likely because the $(2-\delta,0,0)$ position with $\ell=0$ was studied. The present work finds that the scattering is weak for NRXRD measurements for small $\ell$ (see Fig.~\ref{fig:LSCO12_0kl_0q0_map}), this is because the CDW has a large $\mathbf{c}$-axis component to the displacement which reduces the $(\boldsymbol{\varepsilon} \cdot \mathbf{Q})^2$ factor mentioned above.  Refs.~\onlinecite{Christensen2014_CCLF,Thampy2014_TDCI} and the present work only observe strong satellite peaks for $\ell \gtrsim 5.5$. This seems to explain why Ref.~\onlinecite{Wu2012_WBTC} did not observe CDW order with 100~keV x-rays.

For La$_{1.88}$Sr$_{0.12}$CuO$_{4}$, the various studies report transition temperatures in the range $T_{\mathrm{CDW}} = 55 - 85$~K, with similar ordering wavevectors $\delta \approx 0.23$ and in-plane correlation lengths at $T_c$ of $\xi \sim 30-50$~\AA. The large range of CDW transition temperatures is probably due to differences in experimental sensitivity and the range of temperature over which data was collected. In particular, the present work and Ref.~\onlinecite{Thampy2014_TDCI} suggest that there is a component of the CDW correlations that exists up to higher temperatures $\sim 150$~K. The $T_{\mathrm{CDW}}$ being determined from the onset of the stronger low temperature component. Refs.~\onlinecite{Christensen2014_CCLF,Thampy2014_TDCI} observed a suppression of CDW on entering the superconducting state as reported in our data.

Ref.~\onlinecite{Christensen2014_CCLF} also applied magnetic fields up to 10~T and found that the intensity of the CDW satellite peak is field enhanced below $T_c$ demonstrating the competition between CDW order and superconductivity. Ref.~\onlinecite{Thampy2014_TDCI} were also able to show that the $(\delta,0,1.5)$ satellite peak is actually split along $\mathbf{b}^{\star}$ to yield peaks at $(\delta, \pm \varepsilon,1.5)$ with $\varepsilon =0.011$. The resolution in the current paper [e.g. Fig.~\ref{fig:strong_scans_LSCO12}(d)] appears insufficient to resolve the split peaks.

\section{Conclusions}

In this paper we have observed a bulk CDW in three samples of La$_{2-x}$Sr$_{x}$CuO$_{4}$ with $0.11 \leq x \leq 0.13$. While we have not actually studied $x=1/8=0.125$, our data suggests that the CDW is strongest, with the longest correlation length, and highest onset temperature in the vicinity of this hole doping level. The onset temperature of the CDW order ($T_{\mathrm{CDW}}$) is in the temperature range 51--80~K i.e. below the onset temperature of the pseudogap phase in this composition range $T^{\star} \approx 150$~K. $T_{\mathrm{CDW}}$ coincides with long established anomalies in NMR linewidth and the Hall coefficient. The CDW ordering wavevector for $x=0.12$ is (0.235(3),0,0.5).  This is simply related to the wavevector of incommensurate quasi-elastic magnetic order observed by neutron scattering via $\boldsymbol{\delta}_{\mathrm{charge}}= 2\boldsymbol{\delta}_{\mathrm{spin}}$. This contrasts with behavior in YBCO where the strongest low-energy spin fluctuations do not occur at $\frac{1}{2} \boldsymbol{\delta}_{\mathrm{charge}}$.  We find the intensity of the CDW is suppressed on entering the superconducting state, demonstrating strong competition between charge order and superconductivity.  Finally, the temperature dependence of the intensity of the low-energy (quasi-elastic) spin fluctuations appears to track the intensity of the CDW peak. The close relationship between spin and charge correlations in LSCO suggests that the order parameters may be intertwined in real space as in a ``stripe'' pattern.

\section{Acknowledgments}

We thank E. Blackburn, E.M. Forgan and L. Taillefer for discussions. The work was
supported by the UK EPSRC (Grant No. EP/J015423/1).

\bibliography{lsco_I16}
\bibliographystyle{apsrev4-1}

\end{document}